# Shared (Mis)Understandings and the Governance of AI: *A Thematic Analysis of the 2023-2024 Oversight of AI Hearings*

Rachel Leach
Department of Science, Technology, and Society, Virginia Tech

**Abstract**

This paper investigates early legislative deliberations over Artificial Intelligence (AI) in the United States through a thematic analysis of the 2023 - 2024 Oversight of AI hearings held by the Senate Judiciary Committee's subcommittee on Privacy, Technology, and the Law. I focus on these hearings as a site where participants draw from, and renegotiate, accustomed ways of thinking about technology and society. First, I examine how participants, who overwhelmingly represent the technology industry, work to create narratives for understanding the future impacts of AI, and the relationship between technology and progress. Second, I examine how these narratives are invoked to argue for a particular form of AI governance, while casting alternative approaches as everything from infeasible to anti-American. By tracing industry influence over dominant understandings of the impacts of AI and the proper role of government, I examine the arrangements of power enacted and upheld through these hearings. In all, I ask: what role do shared (mis)understandings of AI play in early deliberations over governing this technology?

**Keywords:** *AI Governance, Techno-Solutionism, AI Hype, Legislative Deliberation*

## Background

Artificial Intelligence (AI) is an overarching term which refers to a growing number of technologies that process large amounts of data to complete tasks. An increasing number of American businesses, currently around 78% according to a McKinsey & Company report, use AI for at least one purpose (Hall, 2025). Frequently, AI is marketed towards businesses as a tool that will "mimic the problem-solving and decision-making capabilities of human intelligence," in turn cutting business costs through automation and increased employee productivity (Goodwin, 2025). Companies also market AI towards individuals, including by suggesting that using an AI model will be akin to "chatting with a helpful friend with PhD-level intelligence" (OpenAI, 2025). Despite the expanding prevalence of this technology, its definition is ever-changing (Attard-Frost, 2025).

One notable influence on the spread of the term Artificial Intelligence beyond academic circles is its usefulness in marketing products. As Bender (2023) argues, in many cases, Artificial Intelligence is simply "a marketing term" which companies use to make "certain kinds of automation sound sophisticated, powerful, or magical." In line with this argument, Elish & boyd (2018) find that the terms "Big Data" and "AI" both gained traction in product marketing in the mid-to-late 2010s as companies sought to sell "a future that is scientifically perfectible," through the creation (and purchase of) technologies.

On the other hand, the expanding use of the term could also reflect a belief in the replicable parallels between machines and humans. This vision predates even the use of the term "artificial intelligence," finding its roots in the field of cybernetics and early cultural visions of the internet (Turner, 2006). According to Turner, Norbert Wiener, a leading figure in the field of cybernetics, believed that “biological, mechanical, and information systems” are parallels of each other that can be understood as “forms of pattern and organization" (Turner, 2006, referring to Wiener, 1948). The cybernetic vision of intelligence (both machine and human) is not contextual or subject specific, but all-encompassing. According to R. Buckmaster Fuller, another prominent member of the field, these universal patterns, both social and mechanical, could be understood and directed by a technical expert he called the “comprehensive designer” (Turner, 2006, p. 57 citing Fuller, 1963). For Fuller[1], the role traditional (i.e. non-technocratic) government and bureaucracy should play in this system is minimal, while the role of the technological expert is considerable (Turner, 2006; Fuller, 1981).

The cybernetic vision of the capabilities of machines extended to McCarthy et al. (1955)’s proposed study of “Artificial Intelligence” which was “to proceed on the basis of the conjecture that every aspect of learning or any other feature of intelligence can in principle be so precisely described that a machine can be made to simulate it.” This “conjecture,” operates under nearly identical assumptions as Wiener (1950) who claims, “theoretically, if we could build a machine whose mechanical structures duplicated human physiology, then we could have a machine whose intellectual capabilities duplicate those of human beings.” As Weizenbaum (1976) argues, this vision assumes the strictly “law abiding” nature of computer programs is analogous to the process of human intelligence, without ever proving intelligence operates as such (p. 40). Weizenbaum suggests that this choice to define "man" as merely an “information-processing system” is a reduction required by the logic of the “computer metaphor” (p. 160). This reductionist focus appears in the assumptions of both Wiener (cybernetics) and McCarthy et al. (early “Artificial Intelligence” researchers) and their understanding of “intelligence” (or “intellectual capabilities”).[2]

The term Artificial Intelligence, as it is used now, takes shape both in the context of its usefulness in marketing consumer products, as well as this academic study of the comparable nature of humans and machines. Concerningly, AI as a marketing term and as a field of academic study each reify the power of technological systems and those creating them at times over and above traditional means of democratic governance.

Currently, many of the most advertised and used AI tools, such as OpenAI’s ChatGPT, Microsoft's Copilot, and Anthropic’s Claude are forms of Generative AI (GenAI), a technology which processes large training datasets to generate new content (Hao, 2025).[3] While initially, GenAI was typically confined to responding to human inputs based solely on training data, over time companies have attempted to give these systems more agency. For example, in 2023 OpenAI began to allow certain versions of ChatGPT to access the internet (Radford & Kleinman, 2023). In addition, some companies, including Google and IBM, advertise “AI agents” which they suggest can act on behalf of users with little or no oversight. Google explains that these agents “use AI to pursue goals and complete tasks on behalf of users” and “show reasoning, planning, and memory” (Google, 2025). IBM elaborates on the potential of AI agents explaining that these tools are capable of “decision-making, problem-solving, interacting with external environments and performing actions" independent of human users (IBM, 2024).[4] These increasingly human-out-of-the-loop systems work to transfer control from human users to the machines they are using (and the companies building them); a step towards the goal of many "AI Labs"[5] to develop Artificial General Intelligence (AGI) (Hao, 2025; Metz & Isaac, 2025).

---

[1] For instance, he writes “All the present bureaucracies of political governments, great religious organizations, and all big businesses find that physical success for all humanity would be devastating to the perpetuation of their ongoing activities” (1981). In other words, suggesting that bureaucracies rely on perpetuating suffering.

[2] McCarthy took issue with Weizenbaum’s argument and responds to it here: http://jmc.stanford.edu/artificial-intelligence/reviews/weizenbaum.pdf

[3] For context on just how much data these systems consume, Laizure (2024) explains that “ChatGPT 2 had only 1.5 billion parameters, and the newest version, ChatGPT 4, has 1.76 trillion parameters. The data set used to train ChatGPT 3.5 was 45 terabytes, and the data set for the most recent version (ChatGPT 4) is 1 petabyte (22 times larger than the data set used for ChatGPT 3.5).”

As with AI, AGI does not have a precise definition. Despite this, companies attempting to develop it agree AGI is a machine intelligence that can complete tasks previously only done by humans, especially tasks that are cognitive (according to Google)[6] or economically useful (according to OpenAI).[7] Whether this is possible, or desirable, is still up for debate (Salvaggio, 2025; Heaven, 2025).

Kotliar (2025), explains that the hyperbolic discourse around AI capabilities, exemplified in discussions about AGI, in large part stems from sensational media coverage and advertisements, which together result in "AI hype." Bareis, Roßmann & Bordignon (2023) identify that hype is produced, in part, through the "emotional appeal of overpromising language." Bareis (2026) theorizes a specific form of "hype as exaggerated promises" which, he argues, become influential as they are taken up by the media and culture more broadly. Rather than relying on truth-claims, Bareis (2026) suggests that hype relies on overemphasizing promises of a future where new technology can solve considerable social problems to build support and investment.

Scholars of the sociology of expectation concur that momentum for investment in, and permissive policymaking regarding "newly emergent technologies," typically relies on a "future oriented" focus (Brown & Michael, 2003). For instance, in the context of digital health, Geiger & Gross (2017) find that future-focused language about the possibilities of technologies played a tangible role in producing a market for various forms of digital health care, despite the limited capabilities of existing technologies.

Rather than relying on description and assessment of technologies, technology hype presents a vision of social progress tied to "innovation," technological development, and deployment. For instance, Escobar (1995) identifies that, in the context of "development," economists, primarily from the West, believed that technology would "amplify material progress," as well as produce "a sense of direction and significance" for that progress. As a result, technology was incorporated as a central component of many "development projects" without paying adequate attention to relevant "social and cultural factors" (Escobar, 1995, p. 36).

Bender (2021) argues that, in practice, "AI hype" can be overinclusive by conflating general purpose AI models with tools created for highly specific purposes, such as text-to-speech software, or entirely disconnected from reality, such as the former Google Chief Business Officer's claim that the company is "creating God" (Shah, 2023).[8] Hype works to frame the data intensity and extraction underpinning AI technology as an unfortunate but necessary means to technological, as well as social, progress. Schneider (2024) identifies that this hype empowers industry through the notion that the "regulation of technology is possible only for those who fully understand [it]" (Schneider, 2024).

---

[4] "AI agents" are anthropomorphized through the suggestion that they can perform "reasoning" or "decision-making." This discourse is not new, and instead continues the trend of viewing "algorithms as agential" critiqued by Ziewitz (2016).

[5] The term "AI Labs" can be misleading. Bernath (2025) explains "Many people continue to refer to companies like OpenAI, Anthropic, Google's DeepMind, and Meta's "superalignment" team as "frontier AI labs"." These "labs" developing AI, are fundamentally intertwined with Big Tech companies, and themselves often function to support the development of products rather than scientific research.

[6] Google defines AGI as "a type of artificial intelligence (AI) that aims to mimic the cognitive abilities of the human brain" (Google, 2024).

[7] Altman defines AGI as "highly autonomous systems that outperform humans at most economically valuable work" (OpenAI, 2018).

[8] For many additional examples of "AI Hype", see: Emily M. Bender's "AI Hype Wall of Shame": https://criticalai.org/the-ai-hype-wall-of-shame/ as well as the Hype Studies project: https://hypestudies.org/

While the term AI is highly mutable, and the set of technologies it refers to is frequently expanding, these technologies share a concrete set of common features. AI is a (Big) data driven technology, meaning it relies on large data sets frequently made up of data scraped from the internet, copyrighted works, and sensitive or personal information. Reia et al. (2025) explain, "artificial intelligence (AI) systems depend on collecting and analyzing mass amounts of data, particularly on intimate and private aspects of individuals' lives, identities, and social relations." This data imperative incentivizes companies to collect and make use of data on users, at times without their knowledge or meaningful consent (Waldman, 2021). For instance, in 2024, Meta chose not to give users the option to opt-out of their data being used to train Meta's AI model, Llama, unless they were located in the EU and afforded protection under the GDPR (Weatherbed, 2024).

Foundationally, AI functions by replicating its training material through identifying patterns. As an, albeit sophisticated, probabilistic generation machine, or "stochastic parrot" (Bender et al, 2021), no matter how large AI models get, they do not "know" or "learn," they only predict.[9] Here, a problem arises: AI models answer even simple questions wrong at least some of the time, a phenomenon which is often referred to as "hallucinations"[10] (Li, Yuan, & Zhang, 2024). Even when these machines produce seemingly coherent, and even helpful outputs, Villanueva (2025) cautions us to understand these outputs in context. She explains that AI produces outputs that appear creative or intelligent only by "consuming human history, [and] drawing patterns from our collective creativity." As such, these tools function as a funhouse mirror, reproducing a distorted copy of existing work, existing knowledge, existing creativity, and existing expertise, rather than producing these things[11]

The belief that LLMs, and other consumer products marketed as AI, are knowledge-producing plays an important role in broader visions of the possibilities of these technologies, and the role they ought to play in society. Elish & boyd (2018) explain that the claim that these technologies can, almost magically, perform knowledge-based tasks, serves "as a way to express praise, while also reinforcing a sense that how the technology works is unknowable and inscrutable," This perception, in turn, makes visions of governance and social change through technology all the more compelling, especially to non-experts.

Despite perceptions of automated technologies as disembodied, the material costs of these systems are high (Sobral, 2025).[12] In their paper, *Extractive AI*, Hogan & Lepage-Richer (2024) outline the breadth of environmental harms underpinning the development and spread of AI. They argue that AI relies on extraction ranging from the mining of rare earth minerals to create semiconductor chips to the use of potable water to cool data center servers. Frequently, this extraction affects the most vulnerable disproportionately; for instance, McGovern & Branford (2024) explain AI data centers are frequently cited in Latin America due to "lower environmental regulations than the U.S. and Europe" despite the high risk of droughts through the region.[13]

[9] To be clear, this does not mean LLMs or other GenAI systems cannot produce socially useful outputs (see for instance Esposito, 2022), but rather that those outputs irreducibly rely on vast amounts of preexisting human knowledge and require human interpretation.

[10] Again, this language carries with it particular beliefs about the technology. Specifically, the idea of a "hallucination" frames the generation of incorrect information as a rare fluke rather than a necessary output of this particular kind of technology, along with further anthropomorphizing the technology.

[11] As tech companies continue to build AI models, they also rely on underpaid and precarious labor of marginalized people to moderate and test the technology (Hao, 2025). Roberts (2019) examines the high costs of this labor on the health, economic wellbeing, and communities of the workers performing it.

[12] To be clear, this does not apply to every automated system working to perform useful tasks, but rather to the particular highly data intensive generative models being built without specific, well-defined purposes.

The extractive conditions AI relies on have been repeatedly pointed out to AI companies. For instance, as co-lead of Google's ethical AI team, Timnit Gebru co-authored the paper *On the Dangers of Stochastic Parrots: Can Language Models Be Too Big?* 🦜 (Bender et al., 2021), a project which examined the harms of increasingly large AI models, ranging from bias to expansive energy use. Following this, Google ousted Gebru from the company rather than reckoning with the harms she identified (Hao, 2020).

The development of increasingly large general purpose AI models requires consolidated control over a nearly unfathomable amount of data, infrastructure, energy, and resources across not only the United States, but the Globe (Khanal et al., 2025). Kak et al. (2023) explain that this asymmetrical power means that, "with vanishingly few exceptions, every startup, new entrant, and even AI research lab is dependent on these [Big Tech] firms. All rely on the computing infrastructure of Microsoft, Amazon, and Google to train their systems, and on those same firms' vast consumer market reach to deploy and sell their AI products." For example, OpenAI's products, including the company's most successful consumer facing model ChatGPT, would not exist without the early support and access to infrastructure from Microsoft (Hao, 2025; Naysmith, 2024). So, crucially, the goal of building these "everything machines" (Bender, 2021), and the extraction that underpins this pursuit, is not inevitable, but rather the result of choices made by people in powerful positions at a small number of companies.

**Related Works**

Attard-Frost (2025) argues, "AI governance is frequently ineffective at preventing AI systems from harming society." In this paper, I examine one route through which industry may attempt to render AI governance ineffective: through influencing legislators' understanding of the technology and how to govern it.

This focus is shaped by Winner (1980) who outlines a compelling example of how beliefs about the connection between technology and progress can bias attempts to govern it. Winner explains that in the 1950s, University of California researchers developed a mechanical tomato harvester, which, following a steep initial cost, allowed large scale farms to quickly and cheaply harvest tomatoes with minimal human labor. Following the proliferation of this technology, and the loss of an estimated thirty-two thousand jobs in the tomato industry, an organization representing farm workers harmed by the technology sued the university officials for spending tax money on a technology "that benefit[ed] a handful of private interests to the detriment of farmworkers, small farmers, consumers, and rural California generally" (Winner, 1980, 126). Winner identifies that ascendant understandings of technology at the time, particularly the idea that technologies are inherently neutral tools which cannot, in themselves, cause harm, lead many to view opponents to the automated tomato harvester as "antitechnology" or "antiprogress," preventing the harms they experienced from receiving due consideration and redress (127). I draw from this example a focus on the ways shared understandings of technological progress can lead to outcomes which prioritize technological development over other considerations.

[13] The current and projected environmental impact of AI has been extensively documented including by the UN (UN Environment Programme, 2024). For instance, Bashir et. al (2023) found that "The rapid expansion of Generative AI is reflected in the rising demands on data centers. The datacenter capacity under construction in North America, measured using the datacenter power requirement, increased from 2,688 MW at the end of 2022 to 5,341 MW at the end of 2023 (Beets and Hartnett 2024). This is in addition to the existing demand for datacenters that are expected to add a staggering 12,000 MW of co-location capacity (Uptime 2024), exceeding the demand of over 70% of countries."

A few decades after the development of the mechanical tomato harvester, the public was beginning to use the internet and other novel digital communication technologies. Proponents of these technologies in the United States, including the cyberneticists discussed earlier, attempted to connect them to ideals such as openness, progress, and democracy. Turner (2006) finds that by the end of the 90s, "the libertarian, utopian, populist depiction of the Internet could be heard echoing in the halls of congress, the board rooms of Fortune 500 corporations, the chat rooms of cyberspace, and the kitchens and living rooms of individual American investors." These shared visions of the role and impact of technology, in turn, played a key role guiding in the governance and regulation of these systems. One important output of this is the naturalization of what Kevin Kelly, founder of Wired magazine, called "the New Economy" (Kelly, 1998). Turner (2006) explains that this so-called new economy emerged as traditional worker protections and power eroded under increasing labor fragmentation, in part caused by the spread of the internet. In turn, values centered by many who viewed the internet as liberating, including flexibility and entrepreneurship, helped proponents of the changing economic system obscure its harms.

Utopian understandings of emergent technologies continue to be invoked, including in discussions about the government's role in shaping AI development. For instance, Whittaker (2021) finds that "In March 2020, the National Security Commission on Artificial Intelligence (NSCAI), chaired by former Google CEO Eric Schmidt and helmed by other tech executives, recommended that the U.S. government fund what it termed a national AI research infrastructure, in the name of 'democratizing' access to AI research." Through this process, as with Winner's mechanical tomato harvester, government funds are directed towards research that aligns with the industry's interests in the proliferation of AI, while the funding is framed as in the interest of the public through the discourse of technological development as progress.

Waldman (2021) identifies another discourse pioneered and invoked by tech companies in the face of criticism over their practices of mass data collection and user surveillance. He writes, "The dominant privacy discourse today, from Silicon Valley to Washington, DC, centers around notions of choice, consent, and control; in other words, that privacy is about making our own choices about what to disclose, when, and to whom. It is a vision of privacy so narrow that it allows companies, their employees, and their allies to honestly say they care about privacy and still do little to improve privacy protections for their customers" (6). Waldman argues that through this discourse, and a reliance on vague statements of shared value, companies are able to co-opt and redefine how privacy is understood, in turn continuing their business-model of surveillance and mass data collection. He finds that over time this discourse has been pervasive in influencing governmental understandings of privacy writing, "All in all, in more than fifteen hearings between 2015 and 2020 before the Senate Commerce Committee alone, information industry executives have pushed the discourse of privacy-as-control every time" (68). This understanding was additionally taken up by the FTC whose "reports and guidance documents also frequently defer to and incorporate industry best practices as possible approaches to meeting FTC standards" (110). Based on Waldman (2021), I work to critically consider the potential for generalized shared values to be warped through industry logics and narratives.

Each of the works above highlight the impact value-laden understandings of technological progress generally, and of particular kinds of technologies specifically, have on the governance of these technologies.[14] Building on this foundation, I investigate how AI and technological progress are understood in early legislative discussions on the Oversight of AI in the United States; I am interested in what visions of governance support, and are in turn supported by, these dominant discourses. I chose to focus on these early discussions because, as Winner (1980) explains:

> "By far the greatest latitude of choice exists the very first time a particular instrument, system, or technique is introduced. Because choices tend to become strongly fixed in material equipment, economic investment, and social habit, the original flexibility vanishes for all practical purposes once the initial commitments are made."

My focus on discourse is clarified by Marcuse (1964) who highlights the role it plays in the production of what he calls a "project." He explains that as a discourse becomes a shared means of "comprehending" reality, it "precludes alternative possibilities incompatible with it." Escobar (1995) fathers my understanding, explaining that the creation of a discourse does not presume a "conspiracy" or active coordination. Rather, uniformity is produced through a shared reliance on particular kinds of experts, expertise, and a "context independent" body of knowledge. For the purposes of this paper, my examination of discourse focuses on what I call "shared (mis)understandings", or understandings untethered to an examination of current realities. In other words, I seek to identify instances of arbitrariness behind shared beliefs and highlight the role these assumptions then play in legislators' understandings.

**Methods**

To investigate industry influence over early legislative deliberation, I analyzed a series of six hearings held by the Senate Judiciary Committee's subcommittee on Privacy, Technology, and the Law on the "Oversight of AI". These hearings occurred during the 118th Congress (2023-2024) when Democrats held a majority in the Senate. Each hearing explored a topic relating to the overarching theme of AI oversight including *Rules for Artificial Intelligence; Principles for Regulation; Legislating on Artificial Intelligence; The Future of Journalism; Election Deepfakes; and finally, Insiders' Perspectives*. I accessed both video recordings for each hearing from the Senate Judiciary Committee's official website (see Appendix B for a complete list of hearings and links to transcripts and recordings). In total, across the six hearings, 21 people testified leading to 15 hours of recorded testimony.

To identify the creation of narrative framings throughout these hearings, I analyzed data in three successive stages. First, I created a descriptive table that categorizes each of the individuals who testified at the hearings, to identify who was, and was not, included in these hearings. This table includes participants name, the hearing they participated in, their professional affiliation, and a broader category that represents the organized interest they are a part of including the technology industry, academia, media, government, and civil society (see Appendix A).

I then performed a thematic analysis of the hearings (Lorenzini et al., 2024), first grouping testimony across hearings into broad themes to identify commonly invoked narratives, what McAllum et al. (2019) identifies as "first-level coding". I coded my thematic analysis at the level of the statement, rather than sentence or word because my goal was to identify themes across the testimony broadly, rather than sentence or word level patterns. I identified sections of text ranging from one to six lines, with most passages being about three lines long.

[14] While this paper focuses on industry use of informational asymmetries, control over understandings, assumptions, expertise as a means of shaping AI governance, other factors such as lobbying (Cook, 2024) and the funding of academic research (Burrell and Metcalf, 2024) also play an important role in industry influence over regulatory discussions.

My coding was guided by Lukes' (2005) focus on the often-hidden power enacted through shaping perceived interests and Winner's (1980) discussion of the influence broad understandings of technology can have on the governance of new technologies as they emerge. I sought to identify the dominant shared understandings of AI in these hearings, focusing on those that rely on existing visions of technological progress (Winner) or industry-dominated framings of the public interest (Lukes). To explore this, throughout my coding process I wrote a series of memos organized around the following question: *What are the shared understandings of Artificial Intelligence that arise in the oversight of AI hearings?* I then traced how these shared understandings were deployed to argue for, or against, various governance mechanisms. In particular, I sought to identify and examine the "deployment of discourse" (Escobar, 1995) as a means of producing a specific techno-social vision of the future as well as a shared vision of AI governance in the hearings. I pursue this focus by asking: *What role do shared (mis)understandings of AI play in early deliberations over governing this technology?*

## What are the shared understandings of Artificial Intelligence that arise in the Oversight of AI Hearings?

While coding my data, it became clear that three understandings of AI are frequently invoked by representatives of the technology industry and taken up by Senators. In this section, I present and discuss representative examples of each understanding.

### *Understanding 1: AI has Tremendous Potential Benefits*

Across the hearings, broad and unspecified claims to future possible capabilities of AI are central to the testimony of industry representatives who made up a majority of the participants (see Appendix A). These claims focus on the potential of AI to achieve otherwise unattainable goals in many important areas including health, education, and the environment. For instance, NVIDIA's Senior Vice President William Dalley explains,

*We have made advances in AI that will revolutionize and provide tremendous benefits to society across sectors such as healthcare, medical, research, education, business, cybersecurity, climate, and beyond* (William Dalley, Senior VP NVIDIA, Hearing 3).

In line with the literature on technological hype, frequently industry claims about the benefits of AI involve speculation as to currently unrealized possibilities, rather than existing or concrete examples; OpenAI CEO Sam Altman, for instance, claims:

*We are working to build tools that one day can help us make new discoveries and address some of humanity's biggest challenges, like climate change and curing cancer. Our current systems aren't yet capable of doing these things* (Sam Altman, CEO of OpenAI, Hearing 1).

It is unsurprising that industry representatives claim that their technology, or the technology they will someday create, furthers the public interest; what is notable, however, is that this belief is shared widely across participants. For instance, in an attempt to quantify the benefits of AI, Distinguished Professor of Computer Science at UC Berkeley, Stuart Russell, explains:

*If we succeed [in building AGI], the upside could be enormous. I've estimated a cash value of at least $14 quadrillion for this technology, a huge magnet in the future pulling us forward* (Stuart Russell, Professor UC Berkeley, Hearing 2).[15]

Here, Russell suggests that his predicted technological and economic growth will inherently produce social benefits (as a "magnet ... pulling us forward"). This claim confers upon a not yet existent technology a driving role in the resulting vision of societal progress. Despite this, the arrangements of power that will be produced, or reproduced, by the expanding use of this technology, go entirely unaddressed. Winner's (1980) cautionary tale of the mechanical tomato harvester paints a small picture of the potential upheaval Russell's *magnet* might pull us towards.

Regardless, the potential benefits of this technology are largely accepted by the Senators participating in the hearings. Concerningly, Senators' understandings often lose the initial elements of uncertainty and qualification present in industry testimony, and instead operate as though these claims have been proven; for instance, the Chair of the Subcommittee, Senator Richard Blumenthal (D-CT) explains:

*And there are enormously productive uses, we haven't really talked about them much. Whether it is curing cancer, treating diseases, some of them mundane, by screening X-rays, or developing new technology that can help stop climate change* (Richard Blumenthal, D-CT, Chair of Judiciary Subcommittee on Privacy Technology and the Law, Hearing 2).

Senator Dick Durban (D-IL), Chair of the Senate Judiciary Committee parallels this understanding stating:

*I've heard of the potential, the positive potential of AI, and it is enormous. You can go through lists of the deployment of technology that would say that an idea you can sketch for a website on a napkin can generate functioning code. Pharmaceutical companies could use the technology to identify new candidates to treat disease. The list goes on and on* (Dick Durban, D-IL, Chair of the Senate Judiciary Committee, Hearing 1).

Then, along with specific claims about the potential functions of AI, various Senators believe that technological development necessarily brings social progress. For instance, Senator Corey Booker (D-NJ) argues that technological development functions as a central driver of prosperity and flourishing:

*there's a famous question, ''If you couldn't control for your race, your gender, where you would land on the planet Earth, at what time in humanity would you want to be born?'' Everyone would say, ''Right now.'' It's still the best time to be alive because of technology, innovation, and everything. And I'm excited about what the future holds* (Cory Booker, D-NJ, Hearing 1).

Booker's question is reminiscent of Bareis, Roßmann & Bordignon (2026) discussion of the use of the "emotional appeal of overpromising language" in producing hype (who could possibly be against working towards "the best time to be alive"?). Here, through an allusion to the veil of ignorance, Booker frames government support of technological development and proliferation as universally beneficial for "everyone"; he does not attempt to articulate how these benefits will be equitable, nor where they will come from (aside from the vague notion of "technology, innovation, and everything").

[15] For context, 14 quadrillion is equivalent to 14,000,000 billion or a little over 31,978 times net worth of the world's wealthiest person Elon Musk. 14 quadrillion is also 3,443 times NVIDIA's market cap, 3,935 times Apple's market cap, and 666 times the combined market cap from each of the eight, trillion-dollar, U.S. tech companies, as of September 2025 (Levy, 2025). Finally, 14 quadrillion is 222 times "the total market capitalization of the U.S. stock market", which (in a generous estimate) was $62.8 trillion as of July 1, 2025 (Siblis Research, 2025). In the hearing, Russell does not clarify the evidence he used to come to this conclusion.

The belief that AI will, in the future, bring tremendous benefits works to justify any potential harm of the technology as ultimately worth it. This focus sidelines the significant current harms of this technology, harms which disproportionately affect marginalized people, whether that be older Americans who are especially vulnerable to crimes such as automated scams (National Council on Aging, 2024) or Black communities who are disproportionately affected by the environmental harms of the technology (Bender et al., 2021). Across testimony, the benefits accrued by AI are largely treated as a universal, social good, without an investigation of the potential for the benefits of this technology to be disproportionally accumulated by a small group (Burrell & Metcalf, 2024).

***Understanding 2: AI is Inevitable***

Industry participants frequently suggest that the development of AI is unavoidable. Based on this, they argue that legislation attempting to limit the spread of this technology is neither possible nor desirable. For instance, the Senior VP of NVIDIA William Dally claims:

"*The AI genie is already out of the bottle. AI algorithms are widely published and available to all. AI software can be transmitted anywhere in the world at the press of a button*" (William Dally, NVIDIA, Hearing 3).

Senator Booker (D-NJ) invokes the same metaphor stating:

"*there's no way to put this genie in the bottle, Globally, it's exploding"* (Cory Booker, D-NJ, Hearing 1).

Along with the obvious implication of irreversibility and inevitability, this metaphorframes AI as both quasi-magical and inscrutable, mirroring the concerns of Elish & boyd (2018). If taken to its logical conclusion, the metaphor implies that AI is highly powerful (a Genie grants wishes) in often unknown ways (a Genie often produces counterintuitive results) and its outputs are unstoppable (the Genie is out of the bottle).

Not only do industry representatives, such as Dally, suggest that it would be impossible to limit or prevent existing technology; some frame future technological developments as unstoppable as well. Sam Altman, CEO of OpenAI, constructs a false dichotomy where the only two options his company has is to either benevolently "*[give] people and our institutions and you all time to come to grips with this technology"* or alternatively to "[go] *off to build a super powerful AI system in secret and then dropping it on the world all at once*" (Sam Altman, OpenAI, Hearing 1).[16] While Altman does not explicitly claim that AI, a technology he has personally played a large role in developing, is inevitable, the assumption is implicit through his assertion that the only choice left to be made is when and how the technology will be released. If successful, this understanding would "[preclude] alternative possibilities incompatible with it," as Marcuse (1960) warns.

[16] Of these falsely limited options, OpenAI increasingly chooses the latter (Hao, 2025).

This logic of inevitability has a direct impact on how Senators in these hearings come to think about their role in regulating AI. For instance, Senator Blumenthal (D-CT) applies his belief in technological inevitability to argue that attempts to pause, or curb, the development or spread of AI are unfeasible. He states:

*And for those who want to pause, and some of the experts have written that we should pause AI development, I don't think it's going to happen. We right now have a gold rush, literally much like the Gold Rush that we had in the Wild West, where, in fact, there are no rules, and everybody's trying to get to the gold without very many law enforcers out there preventing the kinds of crimes that can occur* (Blumenthal, D-CT, Hearing 2).

So, despite Altman's worst fear that he and others developing AI will "*cause significant harm to the world"* which he thinks *"could happen a lot of different ways,"* and Blumenthal's belief that at the current pace we may only be *"a couple of years away"* from *"superhuman AI,"* neither are willing to consider regulation to slow the development of this technology, or limit the power of the companies developing it, as a means of limiting these risks. The logic of inevitability leads Blumenthal to conclude that legislation to slow, or meaningfully shape, the development of AI will not happen. Blumenthal maintains this belief even in the face of his understanding that a variety of experts have called for the kind of action he sees impossible.[17]

In many ways, the assumption of technological inevitability makes the job of legislating easier. If AI is on an unstoppable path, then problems that the technology causes, which would otherwise present considerable policymaking challenges to legislators, are ignored. In the words of Michael Rothschild, former Princeton Professor and proponent of the inevitability of capitalism, "being for or against a natural phenomenon is a waste of time and mental energy" (Turner, 2006). On the other hand, an understanding of the contingencies and power behind the emerging technologies legislators are tasked with regulating would necessitate the hard work of developing meaningful legislation in order to uphold the government's responsibility to protect citizens, particularly vulnerable citizens, from a novel, frequently harmful, and elusively defined technology.

***Understanding 3: AI Nationalism[18]***

Industry representatives and Senators participating in the hearings frequently invoke the importance of regulating AI through the "*American way of putting things in place.*" (Amy Klobuchar, D-MN). This framing ties into a broader understanding of AI Nationalism, a belief which asserts the importance of American technology companies building technologies which promote American culture, values, and interests. According to Rijul Gupta, the CEO DeepMedia AI, allowing companies to achieve this requires a distinctly hands off approach, in contrast to alternatives which, he argues, align with foreign value systems. He states:

---

[17] This group of experts includes a number of people testifying at the Oversight of AI hearings such as Yoshua Bengio and Stuart Russell who signed the Future of Life institutes 2023 open letter calling for "all AI labs to immediately pause for at least 6 months the training of AI systems more powerful than GPT-4" in 2023 (FLI, 2023). The Future of Life Institute has its own limitations, for more on the organization see: https://www.politico.com/news/2024/03/25/a-665m-crypto-war-chest-roils-ai-safety-fight-00148621

[18] Ian Hogarth coined the term "AI Nationalism" in his essay which can be found here: https://www.ianhogarth.com/blog/2018/6/13/ai-nationalism. In this paper, I use this term somewhat differently, not only to refer to the role of AI in accelerating tensions between countries, but also to refer to rhetorical framings which suggest that the technology upholds or aligns with a number of national ideals and values.

*I would like to caution us against taking, adopting the Chinese model approach. There's a really great book by a Columbia law professor, Dr. Bradford, where she outlines the Chinese state-driven approach, the European rights driven approach, and the historical United States market-driven approach. So something needs to be done, but a state-driven approach has serious and significant harm to the public, and I want to caution us against adopting regulations that China has put in place.* (Rijul Gupta, CEO DeepMedia AI, Hearing 5)

This *American* approach to regulation holds "innovation" as a core value. Christina Montgomery, the Chief Privacy and Trust Officer IBM, invokes framing suggesting that Senators should only consider limited or "*precision regulation"* so that *"Congress can mitigate the potential risks of AI without hindering innovation"* (Hearing 1). Altman extends the discursive connection between American values and limited technology regulation in his discussion of small businesses.[19] He argues:

*I think it's important that any new approach, any new law does not stop the innovation from happening with smaller companies, open source models, researchers that are doing work at a smaller scale*. (Sam Altman, CEO OpenAI)[20]

Here, Altman presents an updated vision of the American dream: small open-source AI companies thriving in the free market. Protecting this dream, then, relates to a broader understanding of the ideal relationship between the United States and the rest of the world. Industry representatives advance this discourse through the claim that competition and hostility between countries are inherent as each seeks hegemony for their markets and values.[21] The following argument by William Dalley, the Senior VP at NVIDIA, provides a clear example of this:

*So we have to be very careful to balance, you know, the national security considerations and the abusive technology considerations against preserving the U.S. lead in this technology area.* (William Dalley, NVIDIA, Hearing 1).

After being laid out by industry representatives, the connection between unfettered development of AI and protecting American values is taken up by a number of Senators. For instance, Chris Coons (D-DE) suggests that AI reinforced with American values, especially free and open markets, could function as a necessary precaution against the spread of the "*Chinese system.*" He explains:

*The Chinese are insisting that AI, as being developed in China, reinforce the core values of the Chinese Communist Party and the Chinese system. And I'm concerned about how we promote AI that reinforces and strengthens open markets, open societies, and democracy* (Chris Coons, D-DE, Hearing 1)

[19] For an in-depth discussion of the ways the rhetoric of small businesses is used to advance the interests of large, frequently monopolistic companies (see: https://citationsneeded.libsyn.com/episode-111-how-small-business-rhetoric-is-used-to-protect-corporate-america)

[20] And, while Big Tech companies have never been in the business of protecting smaller companies (American Economic Liberties Project, 2025), this framing is especially contradictory as it relates to AI, a space the most powerful tech companies not only acquire AI startups, but also poach founders or workforces of these companies when acquisitions are not viable (Bosa & Wu, 2024).

[21] This framing pulls heavily from the realist theory of international relations which Mirza (2018) explains is not new, and in fact was a central force guiding, and in turn justifying, American foreign policy during the Cold War.

Not only, as AI Nationalism suggests, does American AI provide a necessary counter to the "*Chinese Communist Party,*" it is also potentially protected as freedom of expression, a constitutional right. Senator Blumenthal (D-CT) explains:

*We welcome free expression and AI is a form of expression*[22] *whether we regard it as free or not and whether it's generated and high risk or simply touching up some of the background in the TV ad.* (Senator Blumenthal (D-CT), Hearing 3)

In all, through AI Nationalism, the technology is understood as intertwined with core parts of the American ethos, from innovation and business to protecting democracy. Further, American values, as imbued in the technologies American companies build, are seen as a necessary counterweight to malicious actors, particularly hostile (potentially *communist*) countries developing their own technologies. This narrative connects unfettered American technological development and proliferation to core American values, or ways of " *putting things in place.*" This discourse, then, functions as a means of foreclosing opportunities for critique and meaningful governance.

**What role do shared (mis)understandings of AI play in early deliberations over governing this technology?**

The Senators participating in the hearings express an unquestioned belief in the ability for technology to solve problems. As Senator Booker argues, today is better than yesterday and tomorrow will be better than today "*because of technology, innovation, and everything*" (Hearing 1). Turner (2006) traces this faith in technology as a solution to political and social problems back to the military research world, out of which came the internet, cybernetics, and eventually the study and development of artificial intelligence (Turner, 2006; Bender & Hanna, 2025). Turner explains that central to this belief system is the idea that "experimentation and the proper deployment of the right technologies [can] save the world".

Science and Technology Studies (STS) and information studies scholars call this vision, as it applies to technology governance, "techno-solutionism" (Kneese, 2023; Byrum & Benjamin, 2022; Winner, 1980). Turner's description highlights a few key elements of this form of governance. First, "experimentation" shifts the focus from the current abilities of technology to their future potential. Then, the idea of implementing the "right technologies," negates consideration of the harms of any specific technology, as it was simply not the "right" one or not deployed properly. Finally, the pursuit of "sav[ing] the world" highlights the stakes this vision assigns to continued technological development. As Sætra & Selinger (2024) point out, the centering of technical solutions above alternatives works to frame social problems as "engineering challenges", empowering what Burrell & Metcalf (2024) identify as "a small group of powerful experts," (not dissimilar to Fuller's comprehensive designer) to solve them.

Across the hearings, techno-solutionism is frequently invoked as a solution both to problems created by emerging technologies, and to broader social problems. For instance, Rijul Gupta, CEO of DeepMind AI explains this understanding, testifying that "*AI must be fought with AI" (*Hearing 5). Senators adopt this vision and go so far as to suggest that technical expertise, or a prominent position at a tech company, confers a unique ability to speak on what regulatory solutions ought to be pursued and how. Senator Blumenthal (D-CT), for instance, explains that Zohaib Ahmed, CEO and founder of Resemble AI a voice cloning platform is "*uniquely positioned to understand both the remarkable potential and possible risks associated with the rapid advancement of voice synthesis and cloning capabilities*" (Hearing 5). The sentiment, in turn, shapes proposed legislation. Blumenthal explains "*I can go down our [Bipartisan AI] Framework and tie the proposals to specific comments made by Sam Altman*" (Hearing 3). This vision of governance persists despite the current limitations of technology. For instance, despite invoking testing and the technical implementation of values into AI systems as central mechanisms for maintaining the reliability and safety of these systems, Dario Amodei, CEO of Anthropic, explains, "*we should recognize that the science of testing and auditing for AI systems is in its infancy. It is not currently easy to detect all the bad behaviors an AI system is capable of without first broadly deploying it to users, which is what creates the risk*" (Hearing 2).

[22] This framing of free speech continues in the legacy of *Citizens United v. Federal Election Commission* (https://www.oyez.org/cases/2008/08-205) in empowering corporations to the detriment of the public.

In the hearings, the shared understandings I've identified serve two important functions. First, they allow industry representatives to invoke a vision of AI governance rooted in techno-solutionism without providing evidence that this would be possible given existing technological capabilities. Second, they are used to sidestep or foreclose criticism of existing technologies and practices.

Understanding 1: *AI has Tremendous Potential Benefits,* provides a discursive foundation for the ambiguous sentiment that government intervention comes at a significant risk. As Senator Blumenthal puts it, "*there's also legitimate concern about regulation getting in the way of things, being too cumbersome, and being a negative influence*" (Hearing 1). The "*things*" regulation would get in the way of is left unspecified beyond a vague notion of technology enabled progress, while the "*legitimate concern*" he references is rooted in a future-oriented belief about the ability for future technologies to provide immense social benefit, or as Turner (2006) puts it, "save the world." Crucially, these concerns rely entirely on hypothetical future technological possibilities. The fact that current uses of existing technologies do not live up to this vision is not sufficient to undermine it; the "right technologies" will eventually be developed and implemented correctly. See the following exchange:

> Rijul Gupta, CEO DeepMind AI: *I can say that we are currently engaged with the Big Tech companies. These companies are deploying our technology to fight against deep fake misinformation, and that's already happening.*
> Sen. Richard Blumenthal (D-CT): *Well, they're not doing a very good job.*
> Rijul Gupta: *I know they need to use it more.*

The material requirements, such as energy, water, rare earth minerals, land, and so on, for continued technological development and use (Hogan & Lepage-Richer, 2024), are not addressed. Rather, progress through technology is seen as an almost self-motivating force for progress. This belief makes up Understanding 2: *AI is Inevitable.* This perceived inevitability is at the forefront of Senators minds, with Senator Durbin explaining, "*We're going to have to scramble to keep up with the pace of innovation in terms of our Government, public response to it*" (Hearing 1). This discourse enables the sidelining of the few critical voices at the hearings. For instance, when Woodrow Hartzog, Law Professor at Boston University, suggests that the current trajectory of technological development enables tech companies to "*exploit our data, our labor, and our precarity,*" Senator Mazie Horino (D-HI) responds "*the conclusion I've drawn is that AI is ubiquitous. Anybody can use AI, it can be used in any endeavor. So when I hear you folks testifying about how we shouldn't be taking half measures, I'm not sure what that means"* (Hearing 3). Here, Horino's belief in "*ubiquitous"* nature of AI, directly informs her skepticism of Hartzog's testimony.

A shared belief in technological inevitability works in tandem with personification of the technology. For instance, to downplay concerns over safety, Microsoft president Brad Smith explains: the "*fundamental safety architecture is going to evolve.*[23] *It's going to get better.*" Here, the idea of evolution works to invoke a system of natural improvement to argue that technological safety will also inevitably develop and improve without governmental intervention.[24] The broader belief that LLMs and other forms of Generative AI are knowledge producing lends further credence to this analogy; just as it is not possible to stop, or control evolution, it is not possible to stop technologies which are developing along a set trajectory (a trajectory that, as asserted by *Understanding 1*, aligns with human development and social flourishing).

Here, however, a contradiction seems to arise. On the one hand, technology is seen as inevitably and unchangingly progressing, and on the other, there is concern that legislative intervention will get in the way of said innovation and technological development. In other words, there is both nothing legislators can do to prevent identified harms, while at the same time any attempt at doing so runs the risk of preventing significant benefits. Understanding 3: *AI Nationalism,* is invoked to resolve this dissonance when it arises. For instance, Altman expresses as a central concern the potential that regulation will "*slow down American industry in such a way that China or somebody else makes faster progress*" (Hearing 1). In line with this argument is Senator Coons insistence that "*we promote AI that reinforces and strengthens open markets, open societies, and democracy."* The overarching vision of techno-solutionism is supported by this added premise as it is not just AI, but American AI, that will, as Russell puts it, function as a "*huge magnet ... pull[ing] us forward*."

Techno-solutionism relies on a simultaneously outsized and ambiguous understanding of how this technology works. For example, Blumenthal (D-CT) speculates that to reducing election misinformation and interference, which he suggests is largely perpetrated by foreign countries including Russia and China using AI, can be combated with American technology. He explains, "*If we can use the technology to flip the model, so to speak, if there's a technology that can be used to apprehend and trace the bad guys, that might be a strong deterrent*." How "we" can go about "flip[ping] the model" is not clear. What is clear; however, is that American technology companies, and the technologies they build, are a needed counterforce for good.

Overarchingly, the suggested technical solutions that arise throughout these hearings each rely on speculative claims to future potentials of technology, claims most frequently made by industry representatives of companies building the technologies in question. The dominant techno-solutionism throughout the hearings forces legislators to rely on these industry representatives, who they view as technical experts, to articulate both the future possibilities of their own products, and the socio-political implications of this.

**Conclusion**

The shared understandings of AI and resulting visions of governance are based on speculative and forward-looking claims. Rather than relying on contestable evidence, they construct a techno-social vision in which the current material harms of AI can go largely unconsidered, as at some point in the future they will be vastly outweighed by the entirely novel (and unproven) capabilities of the technology.

[23] When it comes to accountability for existing AI systems, on the other hand, industry representatives criticize such anthropomorphizing language. For instance, Altman explains "*I think it's important to understand and think about GPT–4 as a tool, not a creature, which is easy to get confused*" (Hearing 1).

[24] I would be remiss to not point out the similarity between this line of thinking and Norbert Wiener's belief in the parallels between biological and mechanical systems.

When taken together, these understandings are rife with contradictions. Despite never coming to a firm definition of AI, participants suggest it will contribute to "curing cancer" (Altman) and "help stop climate change" (Blumenthal) all while being worth "a cash value of at least $14 quadrillion" (Russell). At the same time as industry representatives contend the technology is inevitable, they also suggest it is fragile enough to require highly limited and permissive governance, lest its development be stagnated.

As Bender & Hanna (2025) highlight, these "extraordinary claims" ought to require "extraordinary evidence"; yet, throughout the Oversight of AI hearings, industry deployed shared (mis)understandings function as a proxy for this evidence. In turn, many existing harms of the technology, ranging from expanding energy use and environmental degradation (Luccioni et al., 2022; Hogan & Lepage-Richer, 2024) to discrimination and other faulty outputs (Buolamwini, 2023; Bender & Hanna, 2025), are obscured or ignored.

Let's briefly imagine what alternative understandings might look like. Instead of viewing AI as a silver bullet solution, we could commit to an understanding which acknowledges that no technology, on its own, can solve complex social, political, medical, and environmental problems. Instead of resigning ourselves to an inevitable path of technological progression, and the arrangements of power this progression aligns with, we can keep at the forefront of our minds the potential for democratic oversight, organizing, and advocacy. Finally, rather than blindly empowering technology companies to peruse a narrow vision of American nationalism, we can ask weather a technology which is at its root extractive, is even capable of supporting human flourishing in the first place.

**Limitations**

The findings presented in this paper are limited to the specific hearings my data consist of. While the findings align with, and elaborate on, existing theories, they may have limited relevance to contexts beyond the scope of the paper. Further work in exploring subsequent hearings, as well as relevant proposed legislation, would be helpful in examining the breadth and impact of the shared understandings identified in this work.

**Appendix A: Participants, Oversight of AI Hearings**

| Hearing | Name | Affiliation | Category |
|---|---|---|---|
| Hearing 1: Rules for Artificial Intelligence | Samuel Altman | Chief Executive Officer, OpenAI | Tech Industry |
| Hearing 1: Rules for Artificial Intelligence | Gary Marcus | Professor Emeritus, New York University | Academia, Tech Industry |
| Hearing 1: Rules for Artificial Intelligence | Christina Montgomery | Chief Privacy and Trust Officer, IBM | Tech Industry |
| Hearing 2: Principles for Regulation | Dario Amodei | Chief Executive Officer, Anthropic | Tech industry |
| Hearing 2: Principles for Regulation | Yoshua Bengio | Quebec AI Institute (Mila) & University of Montreal | Academia, Tech Industry |
| Hearing 2: Principles for Regulation | Stuart Russell | Professor of Computer Science, University of California, Berke-<br>ley | Academia |
| Hearing 3: Legislating on Artificial Intelligence | William Dally | Chief Scientist and Senior Vice President of Research<br>NVIDIA Corporation | Tech Industry |
| Hearing 3: Legislating on Artificial Intelligence | Brad Smith | Vice Chair and President, Microsoft Corporation | Tech Industry |
| Hearing 3: Legislating on Artificial Intelligence | Woodrow Hartzog | Professor of<br>Law, Boston University School of Law and Fellow | Academia |
| Hearing 4: The Future of Journalism | Danielle Coffey | President and CEO, News Media Alliance | Media Trade Association |
| Hearing 4: The Future of Journalism | Jeff Jarvis | Professor at CUNY Graduate School of Journalism | Academia |
| Hearing 4: The Future of Journalism | Curtis LeGeyt | President and CEO, National Association of Broadcasters | Media Trade Association, Lobby Group |
| Hearing 4: The Future of Journalism | Roger Lynch | CEO, Condé Nast | Media Industry |
| Hearing 5: Election Deepfakes | David Scanlan | Secretary of State, State of New Hampshire | Government |
| Hearing 5: Election Deepfakes | Zohaib Ahmed | CEO and Co-Founder, Resemble AI | Tech Industry |
| Hearing 5: Election Deepfakes | Ben Colman | CEO and Co-Founder Reality Defender | Tech Industry |
| Hearing 5: Election Deepfakes | Rijul Gupta | CEO, Deep Media | Tech Industry |
| Hearing 6: Insiders' Perspectives | Helen Toner | Director of Strategy and Foundational Research Grants, Center for Security and Emerging Technology, Georgetown University | Academia, Formerly Tech Industry |
| Hearing 6: Insiders' Perspectives | Margaret Mitchell | Former Staff Research Scientist, Google AI and chief ethics scientist at AI startup Hugging Face | AI Ethics Research, Tech Industry |
| Hearing 6: Insiders' Perspectives | William Saunders | Former Member of Technical Staff, OpenAI | Formerly Tech Industry |
| Hearing 6: Insiders' Perspectives | David Evan Harris | Senior Policy Advisor, California Initiative for Technology and Democracy, Chancellor's Public Scholar, UC Berkeley | Academia, Formerly Civil Society |

## Appendix B: List of Hearings and Links

***Hearing 1: Rules for Artificial Intelligence***

*Recording*: https://www.judiciary.senate.gov/committee-activity/hearings/oversight-of-ai-rules-for-artificial-intelligence

*Transcript*: https://www.congress.gov/event/118th-congress/senate-event/LC71543/text

***Hearing 2: Principles for Regulation***

*Recording*: https://www.judiciary.senate.gov/committee-activity/hearings/oversight-of-ai-principles-for-regulation

*Transcript*: https://www.govinfo.gov/content/pkg/CHRG-118shrg53503/html/CHRG-118shrg53503.htm

***Hearing 3: Legislating on Artificial Intelligence***

*Recording*: https://www.judiciary.senate.gov/committee-activity/hearings/oversight-of-ai-legislating-on-artificial-intelligence

*Transcript*: https://www.techpolicy.press/transcript-us-senate-judiciary-hearing-on-oversight-of-a-i/

***Hearing 4: The Future of Journalism***

*Recording*: https://www.judiciary.senate.gov/committee-activity/hearings/oversight-of-ai-the-future-of-journalism

*Transcript*: https://www.congress.gov/event/118th-congress/senate-event/LC73543/text

***Hearing 5: Election Deepfakes***

*Recording*: https://www.judiciary.senate.gov/committee-activity/hearings/oversight-of-ai-election-deepfakes

*Transcript*: https://www.techpolicy.press/transcript-us-senate-hearing-on-oversight-of-ai-election-deepfakes/

***Hearing 6: Insiders' Perspectives***

*Recording*: https://www.judiciary.senate.gov/committee-activity/hearings/oversight-of-ai-insiders-perspectives

*Transcript*: https://www.techpolicy.press/transcript-senate-judiciary-subcommittee-hosts-hearing-on-oversight-of-ai-insiders-perspectives/